\documentclass[10pt]{article}

\usepackage{sbc-template}
\usepackage{graphicx,url}
\usepackage[utf8]{inputenc}  
\usepackage{color}
\usepackage{listings}
\usepackage{float}
\usepackage{flushend}

\begin{document}

\title
{
  Towards In-transit Analysis on Supercomputing Environments
  \thanks{The authors gratefully acknowledge CNPq for funding this research.}
}

\author
{
  Allan Santos\and Hermano Lustosa\and Fabio Porto\and Bruno Schulze
}

\renewcommand{\instnum}{}

\address
{
  National Laboratory for Scientific Computing (LNCC)\\
  25.651-075 -- Petrópolis -- RJ -- Brazil
  \email
  {
    \centerline{\{allanmms,hermano,fporto,schulze\}@lncc.br}
  }
}

\maketitle


\newcommand{\lnccInit}{LNCC}
\newcommand{\lncc}{Laboratório Nacional de Computação Científica}

\newcommand{\dexlInit}{DEXL}
\newcommand{\dexl}{Extreme Data Lab}

\newcommand{\comcidisInit}{ComCiDis}
\newcommand{\comcidis}{Computação Científica Distribuída}

\newcommand{\ib}{\textit{Infiniband}}
\newcommand{\iw}{\textit{iWARP}}
\newcommand{\siw}{\textit{Soft-iWARP}}

\newcommand{\petrus}{Petrus}
\newcommand{\virtualis}{VirtualIS}
\newcommand{\sdumont}{Santos Dumont}

\newcommand{\hpcfore}{\textit{HPC4Energy}}
\newcommand{\savime}{\texttt{SAVIME}}

\newcommand{\scholarshipDuration}
{
  Outubro de 2017 a Dezembro de 2017
}


\begin{abstract} 
  The drive towards exascale computing is opening an enormous opportunity for
  more realistic and precise simulations of natural phenomena. The process of
  simulation, however, involves not only the numerical computation of
  predictions but also the analysis of results both to evaluate the simulation
  quality and interpret the simulated phenomenon.  In this context, one may
  consider the duality between transaction and analytical processing to be
  repositioned in this new context.  The co-habitation of simulation
  computation and analysis has been named after in situ analysis, whereas the
  separation in different systems considered as in-transit analysis. In this
  paper we focus in the latter model and study the impact of transferring
  varying block size data from the simulation system to the analytical one. We
  use the Remote Direct Memory Access protocol (RDMA) that reduces the
  interference on performance caused by data copies and context switching.  It
  adopts an in-memory data transfer strategy combined with TCP, using the BSD
  sockets API and the Linux splice(2) syscall.  We present a performance
  evaluation with our work and traditional utilities.
\end{abstract}

\section{Introduction} \label{sec:intro}

\begin{figure*}
  \center
  \includegraphics[scale=0.15,angle=-90]{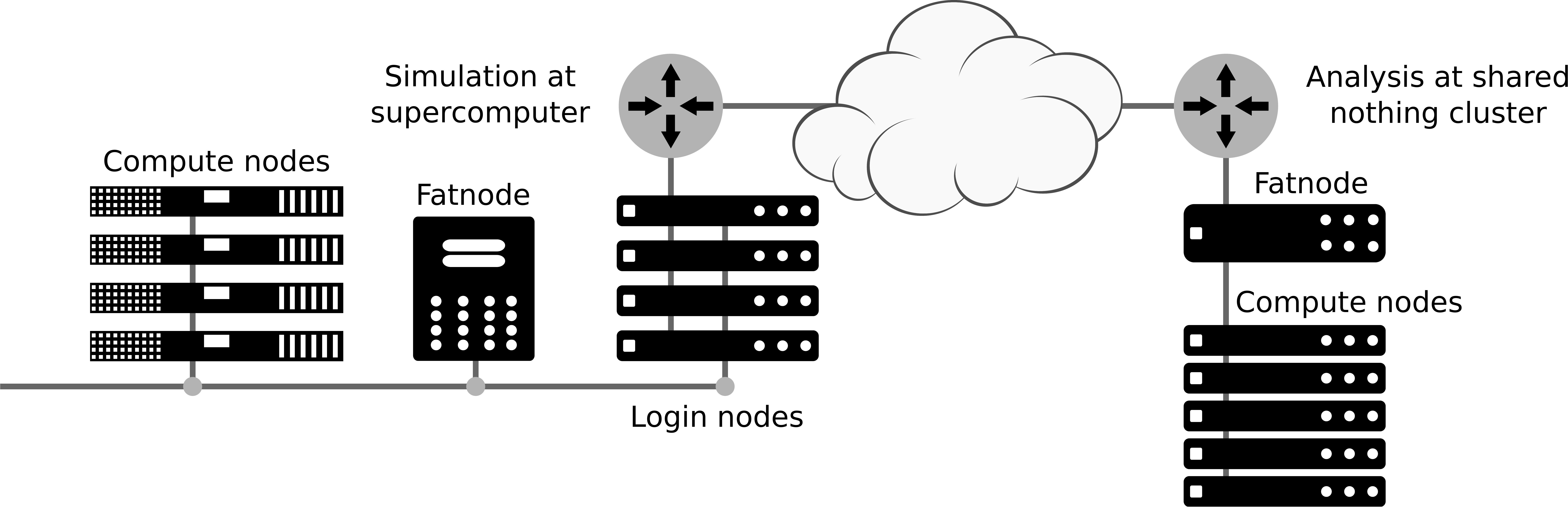}
  \caption{An schema of our target infrastructure.}
  \label{fig:infra-lncc}
\end{figure*}

Computational simulations are becoming more accurate and realistic every day,
leading to more reliable results.  But this also means simulations will
generate larger volumes of data at higher rates.  In traditional,
post-processing, simulation analysis, results are stored on disk before being
accessed by the analytical software.  The whole process is considered slow or
even impractical, when real-time response is expected.  Moreover, traditional
analytical systems, such as database management systems add extra constraints
due to the need of data adjustments between the simulation output and the
expected system data format \cite{bennett2012combining}.

To overcome that, simulation data analytical systems shall be equipped with
tailored data ingestion procedure involving: data representation; data transfer
and assimilation.  The techniques for it are called \textit{in-situ} if
simulation and analytical systems share computational resources.  Otherwise,
they are called \textit{in-transit} \cite{friesen2016situ}.  While the in-situ
analysis allow the process to happen without data transfer, and in some
scenarios without even data copies, the concurrency for CPU and memory can
impose a negative impact on performance of simulation applications.

Conversely, the in-transit model runs the analytical processes at dedicated
environments, avoiding concurrency with simulation for resources.  The
simulation software must be capable to send its results to analysis
asynchronously as they become available.  Similarly, the analytical software
must receive any result as fast as possible and make it available to analytical
applications.

There are some IO frameworks such as NeSSIE \cite{lofstead2011extending}, GLEAN
\cite{vishwanath2011toward} and ADIOS \cite{liu2014hello} can be used to
transfer data from simulation nodes to analytical nodes, which helps to connect
the simulation with an analytical task like visualization as in
\cite{barrett2012report,moreland2011ieee}.  Connecting the simulation with a
database system instead as in \cite{oldfield2009access} allows many simulations
to consume the data and the analysis of past simulations.  But databases
without a data model designed to work with simulation data can impact the
analysis performance.

The computational infrastructure has also an important role in in-transit
analysis.  The network between simulation and analytical environments can have
a significant hop count, i.e., several network devices like routers, firewalls
and proxies in the path.  For this work, we consider two environments at the
same facility but in different networks.  Figure \ref{fig:infra-lncc} shows a
schema of the target computational infrastructure.

Scientific simulation are useful for studying the behavior of natural phenomena
through time.  A mesh is used to define the vertices in space where the
phenomenon occurs or a particular region of interest.  For each time step
previously defined, the mesh is updated with new values for its vertices and
edges.  A native representation for this kind of data are multidimensional
matrices \cite{lustosa2016database}.

In this paper, we present an architectural model and a software library for
staging data on intermediate nodes in the simulation environment while sending
it to a database management system (DBMS) on the analytical environment.  The
DBMS used in this work is an ongoing project at \dexlInit{} at \lnccInit{}
called \savime{}.  It's designed for work with simulation data
\cite{lustosa2017tars} in-memory and has built-in functions for analytical
applications.

Furthermore, this paper explores other characteristics of the infrastructure by
exploiting the available memory resources and high speed networks such as
\ib{}.  To do that, we combine the RDMA protocol \cite{bedeir2010rdma} from
\ib{} with the splice syscall from Linux.  The former reduces CPU consumption
for communication and achieves zero-copy for memory-to-memory data transfer at
the local network.  The later reduces data copies and allows routing traffic
between networks of different environments.

The remainder of this paper is organized as follows.  Section \ref{sec:bg}
presents some technologies and concepts used in this paper.  Section
\ref{sec:arch} describes our software and architectural model and presents some
implementation details.  Section \ref{sec:bench} presents experimental
evaluation.  Section \ref{sec:rel} discusses related work.  Finally, Section
\ref{sec:end} concludes and presents some future works, followed by references.

\section{Background} \label{sec:bg}

%
%

\subsection*{Simulation and Analysis} \label{sec:bg:sim}

Simulation is a process that reproduces natural phenomena by running a
computational model derived from a set of differential equations modeling the
physics of the phenomenon.  The simulation adopts a spatial representation of
the modeled phenomenon in the form of a topology mesh over which values of
predicted quantities of interest are computed.

While running the simulation, scientists want to follow-up on its progress,
which entails accessing and analyzing the simulation output using both error
models and visualization tools.  The former shows how close the simulation is
to observations whereas the latter enables a visual perception of the
simulation outcome in space and time.

The amount of data produced by a simulation is a function of the mesh scale and
size, and the number of time steps.  In fact, for each time step, values of
quantities of interest are computed on all mesh selected points.  Additionally,
the whole set of values are re-computed at each time-step.  Thus, if one wants
to be able to respond to simulation deviations, it is paramount to run data
analysis as fast as possible.  Ultimately, one may expect data analysis task to
be run in real-time with respect to the simulation computation.

\subsection*{RDMA} \label{sec:bg:rdma}

RDMA (\textit{Remote Direct Memory Access}) is a communication protocol with
more operations than the traditional BSD sockets API but using a more low-level
approach.  RDMA allows three types of I/O operations: \textit{two-sided} like
send/receive, \textit{one-sided} like read/write and atomic like
compare-and-swap.  In this work, atomic operations are avoided because some
RDMA devices handle contention very slowly \cite{kaminsky2016design}.

The RDMA two-sided operations are similar to the BSD socket API.  The sender
only knows the source address of the data whereas the receiver only knows the
destination address.  Thus both processes need to be active for the
communication to happen.

However, for one-sided operations only one of the processes is active, i.e., it
knows both the source and destination addresses.  The other process, the
passive side, has sent its part of the needed information previously, usually,
with a two-sided operation.

The passive process can only control the permissions for read or write on a
memory region.  Two-sided operations can be used for synchronization
\cite{bedeir2013basic}, waiting for a control message from client, indicating
that no more remote operations will be executed for a particular memory region
or set of memory regions.

Each application communicating over RDMA needs at least one QP (Queue Pair) for
starting message operations and a CQ (Completion Queue) for receiving
notifications about them.  A CQ can assigned to just one or both queues of a
QP.  A QP can be one of three different types: RC (Reliable Connected), UC
(Unreliable Connected) and UD (Unreliable Datagram).  RC is the most similar to
TCP whereas UD is the most similar to UDP.

The RC type offers 1 to 1 connection for sending and receiving messages in a
reliable way, i.e., it guarantees that packages will be received in order and
without corruptions. Similarly to TCP, a message operation is complete when an
acknowledge is received from the remote side.  An UC QP also offers an 1 to 1
connection but don't guarantees about packet delivery are made.  A message
operation is complete when the entire message was sent.

Finally, the UD type allows both unicast and multicast communications without
reliability.  Thus adding 1 to many communication for RDMA.  Differently from
the two other types, UD message operations don't split a message into packages
of MTU (Maximum Transmission Unit) size, thus the application itself has to do
it and also a message operation is considered complete when its single package
message is sent.

Usually, RDMA is used in networks with specialized hardware called RNICs
(\textit{RDMA-cabable Network Interface Cards}).  RNICs offer capabilities like
loss-free transport layer, network package processing and integrity
verification.  Thus RNICs allows communication without CPU utilization, data
copies and context switches with the OS.

An alternative for a efficient communication with standard Ethernet Gigabit
networks for Linux-based systems is the \texttt{splice} syscall.  It's
available since Linux kernel version $2.6.17$.  With \texttt{splice} one can
move data between two file descriptors without copying data between user and
system space.  Thus reducing CPU and memory consumption for communication.

Since Linux version $2.6.23$, the \texttt{splice} is used to implement the
\texttt{sendfile}\footnote{\url{https://kernelnewbies.org/Linux_2_6_23}}
syscall, a non-standard syscall implemented in most unices for sending a file
through a TCP socket without data copies.  Web servers, like the Apache
HTTPd\footnote{\url{https://httpd.apache.org/docs/2.0/faq/all_in_one.html}},
use \texttt{sendfile} to serve static content.  \savime{} uses standard TCP for
control operations combined with the \texttt{splice} syscall for sending data.

\subsection*{In-memory Scientific Database} \label{sec:bg:memdb}



\savime{} is a novel in-memory DBMS designed for fast data ingestion and
efficient lookup.  To accomplish this, it uses a data model called TARS (Typed
Array Schema) \cite{lustosa2017tars} for managing multidimensional arrays
extended with mapping functions for supporting sparse arrays, non-integer
dimensions, heterogeneous memory layouts and functional partial dependencies
with respect to dimensions.

\section{Architecture} \label{sec:arch}

\begin{figure*}
  \center
  \includegraphics[scale=0.15]{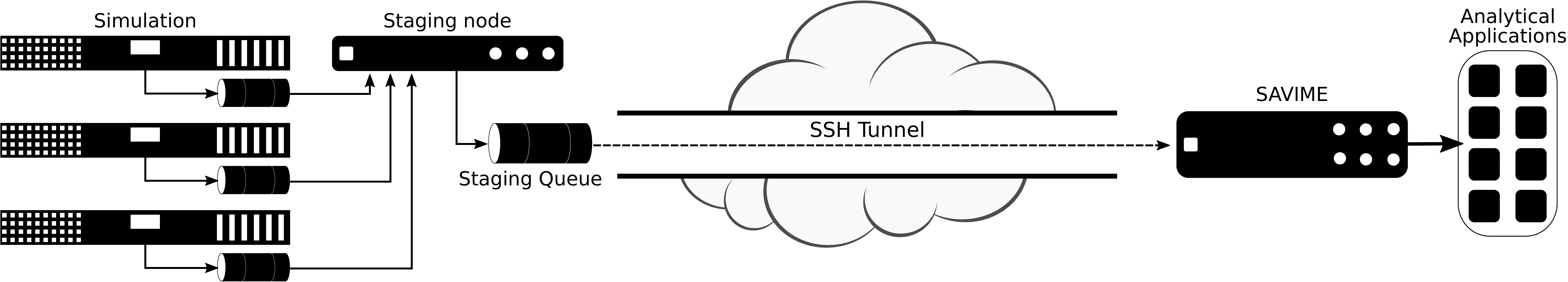}
  \caption{Staging Architecture}
  \label{fig:staging-arch}
\end{figure*}

Nowadays simulations usually write results into distributed parallel file
systems readable from nodes with external network access like a login node.
This is necessary because simulations are executed on machines with access only
to the local cluster network.  The users can then copy the data using utilities
like \texttt{scp} from the intermediate staging node to the analytical
environment.

We make this procedure automatic while reducing disk usage by using the
available main memory and the RDMA protocol in the local fast network.  This
solution has two components: a library (\texttt{libstaging}) and a server
application (\texttt{staging}).  Simulations uses the library to send
asynchronously its data to the server running on staging nodes using RDMA.

\subsection{Overview} \label{sec:arch:overview}

The staging server stores received data as files of a in-memory filesystem
mounted in a previously defined directory.  The size of this mounted filesystem
limits the amount of memory used for staging.  If a file is too big for this
directory (maybe because there is no space left), then the disk is used as a
fallback.  The server sends files in background using TCP/IP to a \savime{}
server on the analytical environment.

The server behaves as a temporary storage.  Received datasets are queued and a
pool of threads sends them in a (First Came First Served) FCFS fashion.
Similarly, the client (\texttt{libstaging}) has a queue of datasets and a pool
of I/O threads sending them to staging.  Thus clients don't block for
communication and can keep working on the next iteration of simulation but the
buffers can't be touched (or released) until completely sent to staging.  Users
can synchronize with the staging server by blocking until all write operations
become finished.

Furthermore, the \texttt{libstaging} can act as a proxy for \savime{} operators
like \texttt{create\_tar} and \texttt{load\_subtar}.  The is needed because
simulation applications can't access \savime{} directly.  These operators
allows the user to describe the structure and organization.  Thus the data can
later be read more efficiently.

Figure \ref{fig:staging-arch} shows a schema of the architecture of
\texttt{libstaging}.  Our first API design allows only one server to be used.
But the staging server creates each dataset independently.  Thus, with a small
modification on the library, clients can send each dataset to a possibly
different server for load balancing.

\lstinputlisting
[
  label=lst:libstaging,
  caption=Sends an arbitrary buffer to \savime{} with \texttt{libstaging}.
]
{
  libstaging.cc
}

Listing \ref{lst:libstaging} shows an example in \texttt{C++} of connecting and
sending data to a staging server running at \texttt{localhost} port 3221.
Lines 2 and 7 shows how to proxy \savime{} commands through
\texttt{libstaging}, omitting its arguments for brevity.  Line 3 creates a
dataset representation using the data type and an user-defined name.  Given an
array \texttt{std::array<double, N>} called \texttt{v}, the
\texttt{staging::dataset::write} method at Line 4 will push a new task to the
local queue.  The \texttt{v} array can't be changed or destroyed until the
\texttt{staging::server::sync} method returns at Line 6.

The development of security measures for staging data is not the goal of this
work; notwithstanding, this concern is considered.  The SSH is a standard
protocol for secure remote shell sessions on the Internet.  But it can also be
used to redirect network traffic between different networks with authentication
and cryptography.  Thus working as a secure tunnel for the data.  In this work,
we use SSH to connect the two environments, i.e., connecting staging server
with \savime{}.

\subsection{Implementation} \label{sec:arch:code}

The current implementation of \texttt{libstaging} and staging server is in the
\texttt{C++} language.  \texttt{libstaging} consists of three main classes
\texttt{server}, \texttt{communicator} and \texttt{dataset}.  The
\texttt{server} holds a \texttt{communicator} object and information about the
staging server like address and port.  The \texttt{communicator} isn't directly
accessed by the final user.  It manages a local queue of tasks and a pool of
threads consuming these tasks as they arrive, using a producer-consumer
strategy.

The \texttt{dataset} object holds a pointer to a memory buffer and a reference
to a communicator.  This reference is used by the \texttt{dataset::write}
method for creating a new task.  Each task is an object with a reference to a
dataset and a method to actually send it to staging.

Once a thread pops a task from the queue, the RDMA communication starts.  We
use a RC QP. It enables the write RDMA operation and allows sending large
messages (up to 2GB) with a single message operation.  After the client
connects to staging server it will send a request for sending the new dataset
informing its name and total size.

The server will \texttt{mmap()} an in-memory file with the dataset size without
touching the mapped memory or registering it for RDMA operations.  Then, the
client starts asking for remote memory blocks it can write to.  Now, the server
register each block as needed before sending the remote memory address
information to the client.

This process is repeated until the client request the memory region for the
last block.  Then, the server posts a receive operation for a synchronization
message.  The client will only send this synchronization message after
completely finish its write operations.  After that, the server can undo the
registration of memory blocks for RDMA operations.

Similarly to \texttt{libstaging}, the staging server has a task queue of
datasets to be sent.  When the last block of a dataset is received through
RDMA, the server will create a new task on the queue.  A poll of threads will
consume these tasks and use the \savime{} client API for create a new dataset
on the database.  Once this operation is complete, the in-memory mapped file
can be removed from the file system to release memory.

\section{Results} \label{sec:bench}

We run experiments with the prototype developed transferring datasets from past
simulations to verify the overhead in time of this library.  For this
preliminary evaluation, we vary block size used in RDMA operations and the
number of I/O threads at each client.  The experiments used the computational
infrastructure of \comcidisInit{} and \dexlInit{} laboratories.

\begin{figure}[htb]
  \center
  \includegraphics[scale=0.75]{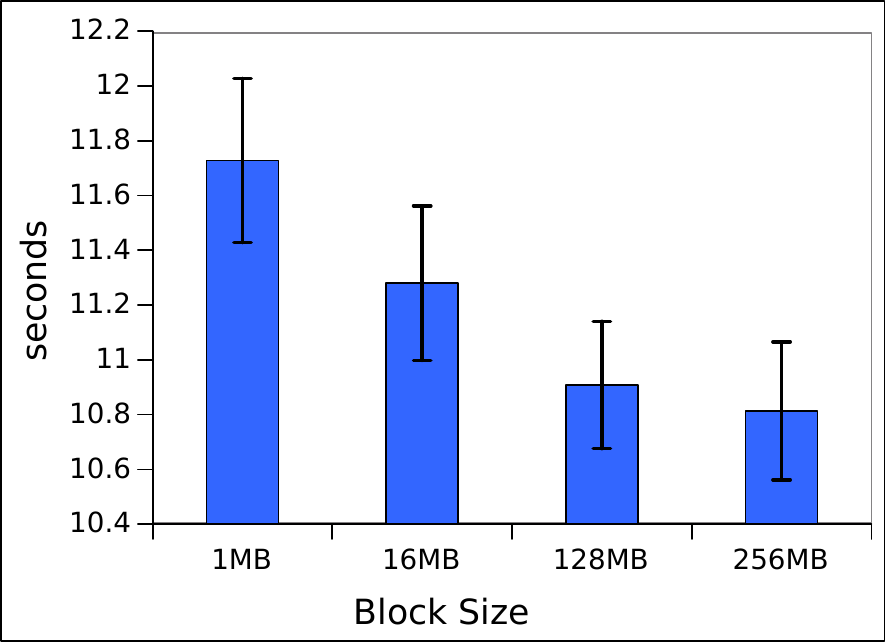}
  \caption
  {
    Elapsed Time (w/ 95\% CI) for sending data to staging with
    \textit{libstaging}, varying block size, using 1 thread per client.
  }
  \label{fig:staging-1t-5x1-95ci}
\end{figure}

The experiment dataset was generated by the HPC4e Seismic Test Suite,
deliverable D(6.3) \cite{puente2015d63}.  The simulation runs 500 trials for a
3D regular mesh with dimensions 201x501x501 containing a velocity field.  The
total dataset have over 25 billion points and more than a 100GB of data.  For
this evaluation, we use a subset of it composed by 85 in-memory files of
approximately 250MB each.  The total size of this subset is 20GB.

The infrastructure used for the experiments presented in this Section consists
of two environments.  The simulation one, from where the data is sent, has 6
identical machines connected by an \ib{} network.  Each machine has 24GB of
memory and 24 CPU cores.  The analytical environment consists of only one node,
a \textit{fatnode}, with 48 CPU cores and 765GB of memory.  This node runs an
instance of the \savime{} DBMS.

\begin{figure}[htb]
  \center
  \includegraphics[scale=0.75]{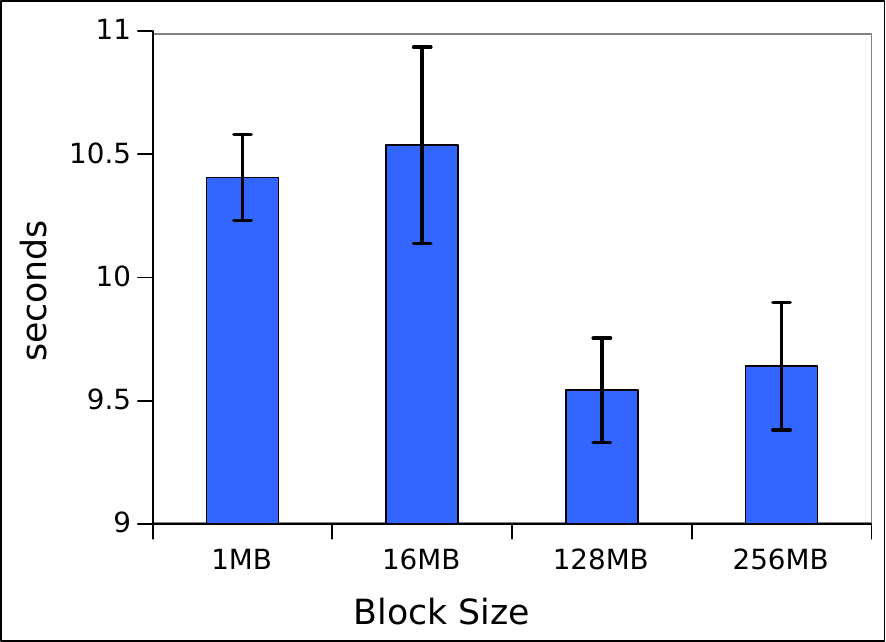}
  \caption
  {
    Elapsed Time (w/ 95\% CI) for sending data to staging with
    \texttt{libstaging}, varying block size, using 4 threads per client.
  }
  \label{fig:staging-4t-5x1-95ci}
\end{figure}

One machine of the simulation environment has an IP interface with access to
external networks.  This machine is the staging node but also the login node of
this cluster.  The other ones are computing nodes.  Each has a subset of the 85
files and send its data to staging.

With this experiment we analyze the time need for transferring data from
computing nodes to the analytical one.  The staging node spent about 3 minutes
and 10 seconds for sending all the dataset via the TCP/IP network.  This time
is constant for this experiment because we only vary the parameters at the
clients. 

Figure \ref{fig:staging-1t-5x1-95ci} shows the elapsed time for transferring
the 20GB dataset varying the block size.  As expected we got better results
with larger blocks.  This figures shows results for only one I/O thread at each
client.  This configuration allows non-blocking communication with the minimum
concurrent for network.

We can reduce elapsed time by using more I/O threads as shown in Figure
\ref{fig:staging-4t-5x1-95ci}.  However, the results become less stable with
the increase on concurrency.  Large blocks also minimizes that.

\begin{figure}[htb]
  \center
  \includegraphics[scale=0.75]{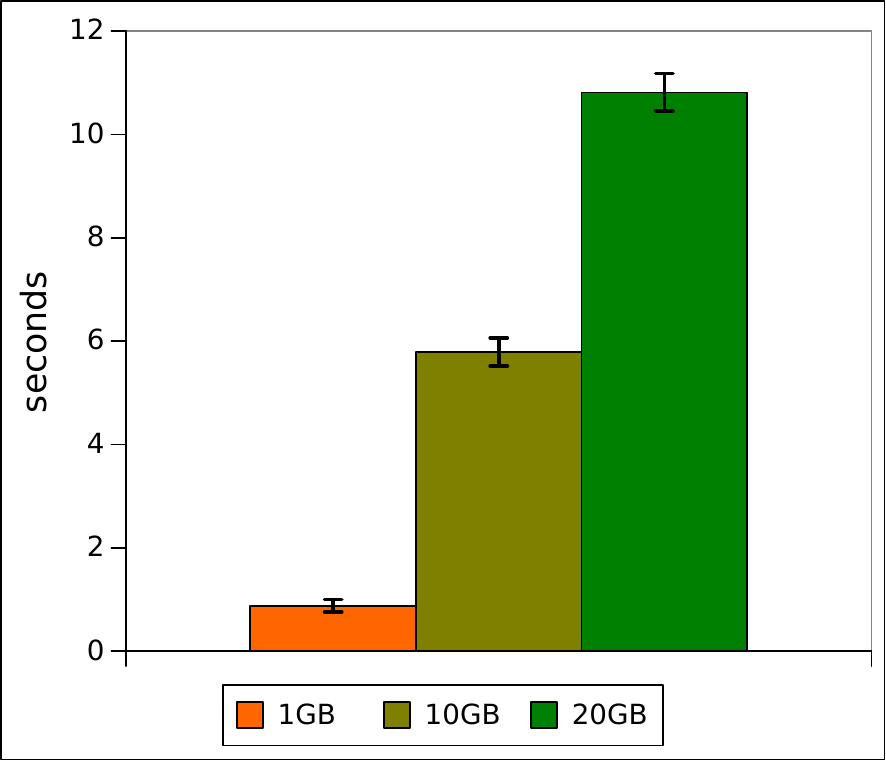}
  \caption
  {
    Elapsed Time (w/ 95\% CI) for sending data to staging with
    \texttt{libstaging}, varying dataset size, using 1 thread per client and a
    fixed maximum block size to 256MB.
  }
  \label{fig:staging-scale-1t-256M-5x1-95ci}
\end{figure}

We also ran \texttt{libstaging} varying the dataset size and fixing the number
threads to 1 and the block size to 256M.  Figure
\ref{fig:staging-scale-1t-256M-5x1-95ci} shows the elapsed time scaling
linearly with the increase on dataset size.


As a baseline method for comparison, we use the standard \texttt{scp} transfer
and the \texttt{pdsh} command to make it parallel.  We use the same
infrastructure, including the \ib{} network and the \savime{} API for sending
data from staging to the analytical environment.  Two types of storage are used
in-memory and disk.

\begin{figure}[htb]
  \center
  \includegraphics[scale=0.75]{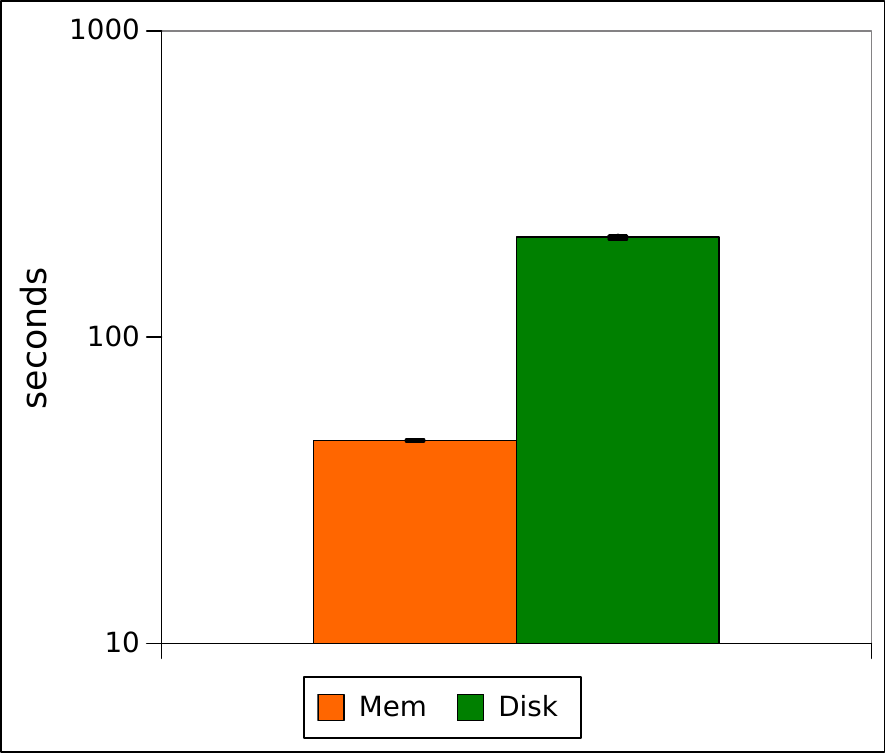}
  \caption
  {
    Elapsed Time (w/ 95\% CI) for sending data to staging with \texttt{scp},
    using \texttt{pdsh} to allow parallel execution.
  }
  \label{fig:no-staging-pdsh-scp-95ci}
\end{figure}

As shown in Figure \ref{fig:no-staging-pdsh-scp-95ci}, for the in-memory
storage, the time spent transferring files from computing nodes to staging is
about 45 seconds, 4 times slower than the RDMA based staging.  The disk storage
adds a huge overhead, increasing the elapsed time to about 3 minutes, 18 times
slower than \texttt{libstaging}.

Another approach is to use the \texttt{SSH} protocol to redirect the \savime{}
commands from the compute nodes directly to the analytical environment.  This
requires two forwards: from compute nodes to staging, and from that to
\savime{} node.  The whole process is transparent for the API.  Even using
in-memory data and TCP over the \ib{} network, the elapsed time is about 4
minutes.

\section{Related Work} \label{sec:rel}

Nessie (NEtwork Scalable Service Interface) \cite{lofstead2011extending} is an
RPC mechanism designed to take advantage of RDMA and related high speed network
technologies for implementing an efficient I/O API, using collective I/O
libraries.  Additionally, applying pre-processing at the staging nodes for data
consolidation reducing the number of I/O operations for accessing data on
storage. 

ADIOS (Adaptable IO System) \cite{liu2014hello} is a io framework to write
files with a custom format, called BP, at staging nodes in parallel using
several backends without need for recompilation.  Users must describe the I/O
operations in a XML file and create ADIOS files, not necessarily regular POSIX
files, in the program.
 
Optionally, a script provided by ADIOS translate the I/O operations to C code
which users must include in their program.  The XML file is also used at
runtime.  Users can change the transport methos used to manipulate the files.
Available options include POSIX and MPI-IO standard interfaces.  Another
options includes Nessie.

Although these projects successfully offload the I/O operations to a set of
staging nodes to keep the compute nodes producing data, they are designed to
run simulation and analysis at the same computational environment, i.e., the
same cluster.  The goal of this project is to use a completely independent
environment for analysis.  Moreover, the cost of converting the BP format for
ADIOS can become a bottleneck for I/O operations.

ICEE is a framework \cite{choi2013icee} for in-transit data analysis in
wide-area networks based in ADIOS.  To accomplish this ICEE applies data
reduction by indexing and filtering data at staging.  Despite reducing the
amount of data sent, these capabilities can impose undesirable processing load
at the simulation environment.  It becomes more relevant if staging nodes used
in productions are shared with other users, e.g., login nodes used to compile
programs, submit tasks and transfer source files.

GLEAN is a infrastructure for simulation-time analysis with non-intrusive
integration with applications \cite{vishwanath2011toward}.  It can use the
sockets API to forward data between environments in different networks for
in-transit analysis.  Furthermore, it offers data semantics since the staging
phase by design, allowing analytical procedures to run at staging nodes to
reduce the volume of data at storage.

Our design, by the other hand, keeps the simulation data as multi-dimensional
arrays because it's the native format used by \savime{}, avoiding any
adjustment or conversions to send data as fast as possible.  \savime{} buit-in
functions can be used to reduce data movement at the analytical environment.
Moreover, as a DBMS, \savime{} allow many applications to consume simulation
results concurrently, distributed across the analytical environment.

\section{Conclusion and Future Work} \label{sec:end}

The load imposed by analytical procedures can impact the performance of
numerical simulations.  The use of in-transit analysis can improve this by
running analytical procedures on a dedicated environment.  Moreover, we can
reduce the load on simulation machines from context switches data copies
between system and user space during communication using the RDMA protocol.

The \texttt{libstaging} presented in this work helps simulation developers to
send simulation results as they are generated.  We use the computational
infrastructure of common environments for simulations and supercomputers for
design this library.

As storage we use a novel database management system designed for fast
ingestion of simulation data called \savime{}, so analytical procedures can
consume simulation results.  Finally, we provided performance evaluation by
transferring experimental data from Seismic Test Suite using our
\texttt{libstaging} and other common techniques used for post-processing
analysis like \texttt{scp}.

Data reduction techniques can be applied on staging nodes to speed up
analytical processes by reducing the need for transferring data.  But only the
application consuming simulation results from \savime{} knows the ranges of
interest.  The \savime{} API already allows filtering stored data by dimensions
and by range.

As a future work, we could add data reduction capabilities on staging nodes by
forwarding filter queries from \savime{} to them.  Thus allowing applications
to analyze only a selected range and ignore the remainder at the staging nodes,
saving time and network bandwidth.

\bibliographystyle{sbc}
\bibliography{references}

\end{document}